\documentclass[conference]{IEEEtran}
\IEEEoverridecommandlockouts
\usepackage{cite}
\usepackage{amsmath,amssymb,amsfonts}
\usepackage{algorithmic}
\usepackage{graphicx}
\usepackage{textcomp}
\usepackage{xcolor}
\usepackage{babel}
\def\BibTeX{{\rm B\kern-.05em{\sc i\kern-.025em b}\kern-.08em
    T\kern-.1667em\lower.7ex\hbox{E}\kern-.125emX}}
\begin{document}

\title{Using Mobility Patterns for the Planning of Vehicle-to-Grid Infrastructures that Support Photovoltaics in 
Cities\\}

\author{\IEEEauthorblockN{Markus Schläpfer, Hong Jun Chew, Seanglidet Yean and Bu-Sung Lee}
	\IEEEauthorblockA{School of Computer Science and Engineering, Nanyang Technological University, Singapore \\ $\{$schlaepfer, hchew007, seanglid002, ebslee$\}$@ntu.edu.sg}}

\maketitle

\begin{abstract} 
	The vehicle-to-grid (V2G) concept utilises electric vehicles as distributed energy storage and thus may help to balance out the intermittent availability of renewable energy sources such as photovoltaics. V2G is therefore considered to play an important role for achieving low-carbon energy and transportation systems in cities. However, the adequate planning of city-wide V2G infrastructures requires detailed knowledge of the aggregate mobility patterns of individuals and also needs to keep track with ongoing developments of urban transportation modes. Here, we introduce an initial framework that infers population-wide mobility patterns from anonymised mobile phone location data and subsequently superimposes a vehicle charging and discharging scheme. The framework allows for the estimation of the aggregate V2G energy supply and demand at fine-grained spatial and temporal scales under a given electric vehicle usage scenario. This information provides an adequate basis for assessing the role of V2G in the context of maximising the deployment of photovoltaics, as well as for the sizing and placement of the required vehicle (dis)charging infrastructure. The proposed framework is applied to Singapore as a case study.
\end{abstract}


\section{Introduction}
Tackling the global challenge of climate change requires a rapid technological transition towards low-carbon \mbox{cities \cite{ipcc_2019}}. Propelled by the recent uptake of plug-in electric vehicles (EVs) \cite{muratori_2020} and photovoltaics (PV) \cite{fraunhofer_2021}, V2G is considered as a promising concept because it may contribute to the deep decarbonization of both the urban energy and transportation systems \cite{lund_vre_2015, gschwendtner_v2x_2021}. By allowing EVs to feed the energy stored in their batteries back to the electricity grid (bidirectional power flows), V2G offers potentially large amounts of decentralized energy storage devices. Since vehicles are typically idle most of the time, EVs thereby offer a large time window for battery charging or discharging. This highly flexible energy storage may ease the large-scale integration of renewable, `carbon-free' energy sources such as PV and wind by balancing out their intermittent availability (e.g., by storing energy from PV during the day and feeding energy back to the grid during the night) \cite{lund_vre_2015}. Indeed, a techno-economic analysis for Kyoto has shown that such a city-wide `PV + EV' system could reduce the city's CO$_2$ emissions from vehicle and electricity usage by 60–74\% \cite{kobashi_soci_2020}. At the building scale, using building-integrated PV and a hydrogen fuel cell EV for combined mobility and power generation has been shown to reduce the imported electricity from the grid by $\approx\,$71\% \cite{robledo_evsol_2018}.

Most of these expectations are based on simplified assumptions with respect to the underlying mobility \mbox{patterns \cite{muratori_2020}}. For instance, people are often assumed to (dis)charge their vehicles only at home or at work. Moreover, it is typically assumed that the entire existing fleet of Internal Combustion Engine (ICE) vehicles will be replaced by EVs \cite{kobashi_soci_2020}, which contradicts current efforts to reduce private car usage. More generally, uncertainties regarding the future development of multi-modal urban transportation systems (including disruptive technologies such as autonomous vehicles) are hardly taken into account \cite{muratori_2020}. These uncertainties regarding the future role of V2G are exacerbated by the ongoing COVID-19 pandemic, which is changing the way people move in cities due to shifts towards online shopping and work-from-home.

Keeping track with population-wide mobility patterns at relatively high spatial and temporal resolution is however crucial for the adequate planning of V2G systems and their interplay with renewable energy sources. This is particularly relevant for the optimal placement of renewable-based EV charging stations or for assessing the need to extend local electricity distribution networks or microgrids so as to cope with locally high peak demands \cite{muratori_uec_2018}. To gain such knowledge, the combination of methods from urban analytics, urban science and infrastructure engineering offers a promising approach. Indeed, the use of `urban big data', such as anonymised mobile phone records or data from location-based social networks, has recently proven to provide fine-grained insights into the mobility behaviour and resulting local energy demands \cite{martinezcesena_mep_2015, selvarajoo_ulf_2019, salat_epe_2020, happle_uom_2020}. In particular, the use of mobile phone location records has previously been proposed for a data-driven optimization of EV charging station locations \cite{vazifeh_des_2019}. However, to the authors' knowledge no quantitative study has so far investigated the potential applicability of mobile phone location data for V2G infrastructure planning.

The aim of this paper is to assess the feasibility and benefits of using extensive mobility data for the planning of city-wide V2G systems that support the large-scale deployment of photovoltaics in cities. To that end, we introduce an initial framework that infers the aggregate movements of individuals from anonymised mobile phone location data and superimposes a simple scenario for the fraction of trips covered by EVs. The main assumption is that aggregate movement patterns are not very sensitive to changes in the transportation mode \cite{schlaepfer_uml_2021}. Subsequently, the framework introduces a spatio-temporal charging and discharging scheme that aims to balance out the intermittent availability of solar energy so as to maximise the deployment of PV (through day-time charging and night-time discharging). As a result, the framework allows to quantify the expected V2G energy supply and peak demand at fine-grained spatial and temporal scales that are not restricted to home or work locations. 

The proposed framework is applied to Singapore as a case study. The city-state with a current population size of $n_\text{pop} \! = \! 5.5 \! \cdot \! 10^6$ is following an ambitious sustainable development agenda. It aims to substantially extend its solar energy deployment to 1.5 GW-peak by 2025 and fosters the switch from ICE vehicles to EVs through installing 60'000 charging points by 2030 \cite{gos_sgp_2030}.

\section{Spatial and temporal V2G modelling framework}
\subsection{Human Mobility Data}

The mobility dataset used in this study consists of \mbox{$\approx \, $4.2 million} GPS-based location records per day from \mbox{$\approx \, $6$\cdot$10$^5$} anonymised mobile phone users in Singapore during the month of September in 2020. Each record is characterised by a geographic position (latitude and longitude) and a time stamp. Following the approach in \cite{schlaepfer_uml_2021}, we aggregated the location records into a regular grid with cells of size \mbox{250\,m $\times$ 250\,m}. For each user, we then identified the set of visited locations (grid cells). To do so, we imposed a minimum stay time $\tau \!\!= \,$1h, during which the user had to stay inside a given cell to be counted, removing those cells where the user only travelled through without engaging in some form of activity. This resulted in a time-ordered sequence of visited locations (trajectory) for each user \cite{schlaepfer_uml_2021}. Finally, we considered only regularly active users for whom trajectories were available during at least five consecutive days ($n_\text{usr} \! = \,$7.2$\cdot$10$^4$).

\subsection{Vehicle (Dis)Charging Scheme}

The model for the vehicle (dis)charging behaviour is based on the following assumptions:
\begin{enumerate}
	\item Depending on the battery state-of-charge (SOC) users charge or discharge their vehicles. For simplicity, our model assumes a fixed SOC threshold, $c_\text{thr} \in \left[0,1\right]$. If the actual SOC is below this threshold, the users charge the battery of their vehicles. Conversely, if the SOC is above the threshold, they are able to discharge their vehicle.
	\item To be completely general, our model assumes that users can (dis)charge their vehicles in any location where they stay for a minimum duration of $\tau \!\!= \,$1h. As such, charging and discharging are not tied to the home or work locations. Following \cite{muratori_uec_2018}, we further assume uncoordinated EV charging and discharging, so that users (dis)charge their vehicle immediately (following the additional rules below) and at fixed power levels, $P_\text{charge}$ and $P_\text{discharge}$, respectively. 
	\item To focus on the operational support for photovoltaics, we assume that users charge their vehicles during the day, when the solar irradiance is sufficiently high. Conversely, they discharge their vehicles during the evenings, night hours and early mornings, when the solar irradiance is low. We thus use a fixed time interval for photovoltaic-based vehicle charging, $\left[t_s^\text{start}, t_s^\text{end} \right)$, derived from the daily irradiance profile. Outside of this time interval, the users charge their vehicle from power sources other than photovoltaics if the SOC falls below the threshold $c_\text{thr}$.
	\item Finally, following \cite{vazifeh_des_2019}, we assume that the battery depletion due to driving, $\Delta E_d$, is, on average, a linear function of the distance travelled. It is thus given as $\Delta E_d = c_\text{max} / l_\text{max} \cdot l$, where $c_\text{max}$ is the battery capacity (available energy when SOC$=$1), $l_\text{max}$ is the maximum distance range of the vehicle and $l$ is the actual driven distance between two locations.
\end{enumerate}
Fig.~\ref{fig:v2gmap} illustrates the resulting (dis)charging regimes of an individual user who visits different locations (A$\rightarrow$B$\rightarrow$C$\rightarrow$D$\rightarrow$A) over the course of a day. The different regimes determine how much energy the vehicle provides to the visited locations (discharging regime), and how much energy the visited locations need to provide to the vehicle, either through PV (`PV charging regime') or through other sources (`non-PV charging regime'). Energy losses (e.g., from power electronics such as inverters) are not considered.

\begin{figure}[tbp]
	\centerline{\includegraphics[width=9cm]{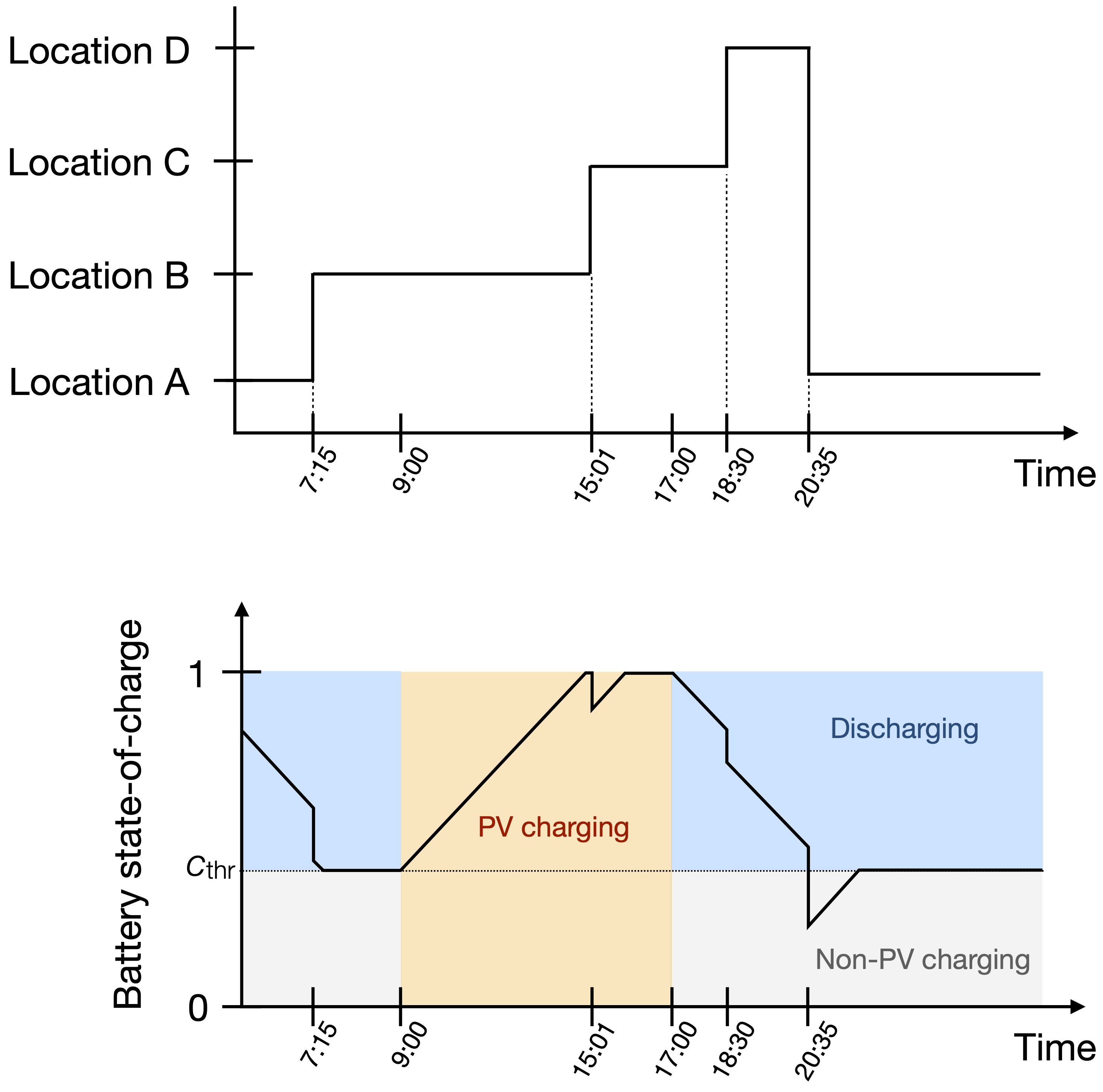}}
	\caption{Illustration of the different EV battery charging and discharging regimes for an individual user. Top panel: trajectory of the user across the visited locations. Bottom panel: resulting battery state-of-charge (SOC) over time. The battery depletion due to driving between the locations is indicated by the vertical jumps in the SOC.}
	\label{fig:schematics}
\end{figure}

\subsection{Aggregated V2G Energy Supply and Peak Demand}

To assess the impact of the considered V2G scheme on the energy supply and demand in different regions of a city, we aggregate the contributions of the individual users (Fig. \ref{fig:schematics}) into geographic areas. The total V2G energy supplied to area $k$ during a given day is estimated as:
\begin{align}
	E^\text{ev}_k \approx \sum_i^n E^\text{ev}_{i,k} \cdot \delta / s
\end{align}
where $E^\text{ev}_{i,k}$ is the net energy input of user $i$, $\delta \in (0,1]$ is the population-wide EV penetration rate and $s = n_\text{usr}/n_\text{pop}$ is the mobile phone coverage of the population (market share), with $n_\text{pop}$ being the city population size. The parameter $\delta$ can be varied according to different EV usage scenarios.

Similarly, to estimate the local PV power generation required to cover the demand under different EV penetration scenarios, we calculate for each area the peak charging demand as:
\begin{align}
P_k^\text{ev,peak}  \approx \max_{t \in T} \sum_{i}^n P^\text{ev}_{i,k,t} \cdot \delta / s
\end{align}
where $P^\text{ev}_{i,k,t} \geq 0$ is the charging demand of user $i$ within area $k$ during time step $t$ within the period $T$ of one day.

\section{Case Study Application}

Singapore consists of 55 planning areas which are the main urban planning divisions. We use these planning areas as the geographic units of analysis, since they provide an adequate spatial aggregation for the planning and operation of integrated multi-energy systems at the urban district-scale \cite{mancarella_2014, fonseca_2015}. The parameter values of the V2G model used for the case study are shown in table \ref{table:parameters}. For each day, the simulation of the V2G scheme runs midnight-to-midnight. The SOC at the start of each simulation is set to a fixed value of $c(t_0) = 0.5$ for all vehicles.
\begin{table}[tbp]
	\caption{Parameters of the V2G model}
	\begin{center}
		\begin{tabular}{|c|c|c|c|}
			\hline
			\textbf{Parameter}&\multicolumn{1}{|c|}{\textbf{Value}}  & \textbf{Parameter}&\multicolumn{1}{|c|}{\textbf{Value}}\\			
			\hline
			$c_\text{max}$&  25\,kWh &  $l_\text{max}$ &  135\,km  \\
			$P_\text{charge}$&  6.6\,kW  & $c_\text{thr}$ &  0.5  \\
			$P_\text{discharge}$&  6.6\,kW & &  \\
			\hline
		\end{tabular}
		\label{table:parameters}
	\end{center}
\end{table}
Based on the empirical irradiance distributions presented in \cite{khoo_opv_2014}, we set the daily time window for the PV charging as $t_s^\text{start}\!\!=\,$9am and $t_s^\text{end}\!\! =\,$5pm.

\begin{figure}[t!]
\centerline{\includegraphics[width=9cm]{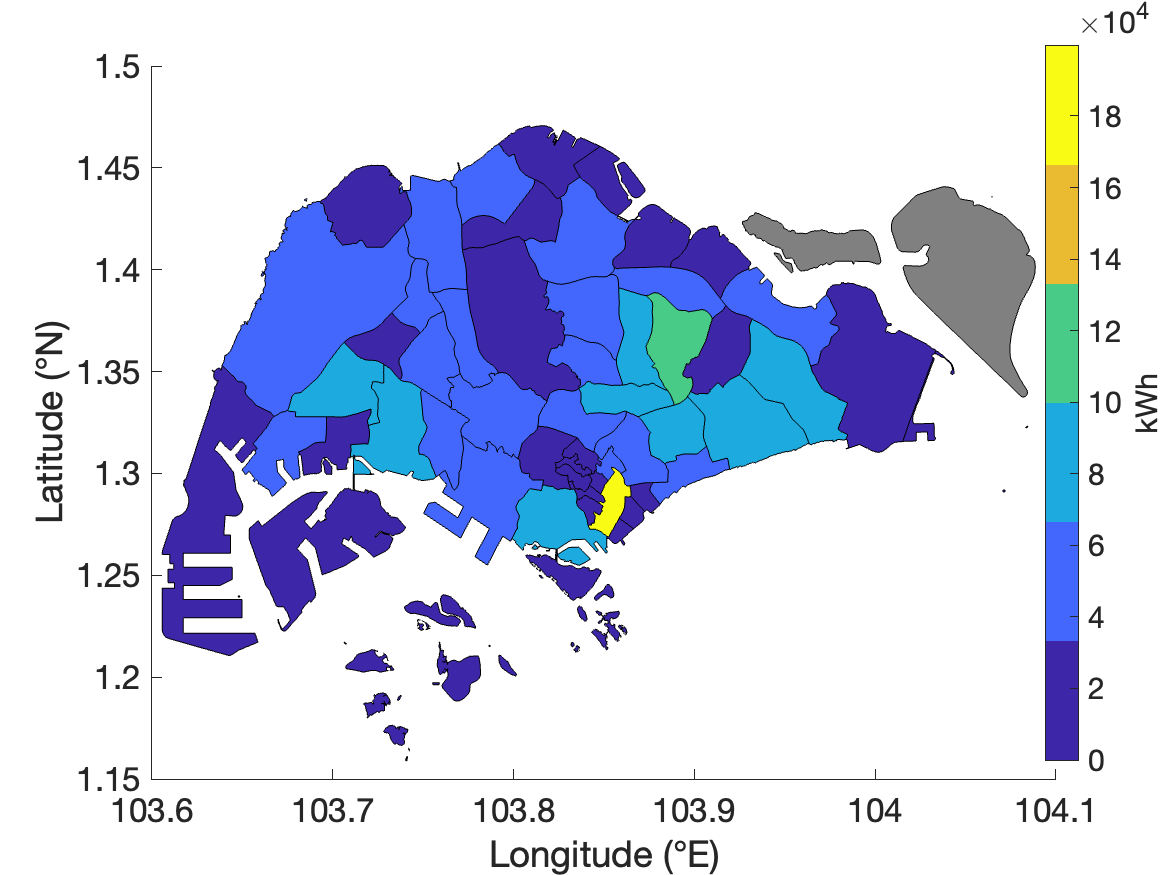}}
\caption{Estimated local energy supply, $E^\text{ev}_k$, through V2G in each planning area of Singapore for a typical day. Values correspond to a low EV penetration rate of 3\% ($\delta = 0.03$). The largest values of $E_k$ are found in the downtown area ('downtown core') as well as in the town centres on the east and west coast.}
\label{fig:v2gmap}
\end{figure}
Fig.~\ref{fig:v2gmap} shows the resulting estimations of the energy, $E^\text{ev}_k$, provided to each planning area $k$ by the described V2G scheme. The values of $E^\text{ev}_k$ vary strongly across the different areas. Importantly, we observe that areas with large amount of amenities (e.g., shopping malls) tend to receive particularly high energy inputs. A possible explanation is the fact that people tend to spend time in these areas for different early-evening activities (e.g., for shopping or having dinner), before returning to their home location. During those activities the EVs are in the discharging regime (see Fig.~\ref{fig:schematics}). As a result, and in contrast to common assumption \cite{lazzeroni_2019}, a large part ot the energy from V2G can be potentially supplied to areas outside residential neighbourhoods (under the premise of the simple, uncoordinated V2G scheme described above).

\begin{figure}[t!]
	\centerline{\includegraphics[width=9cm]{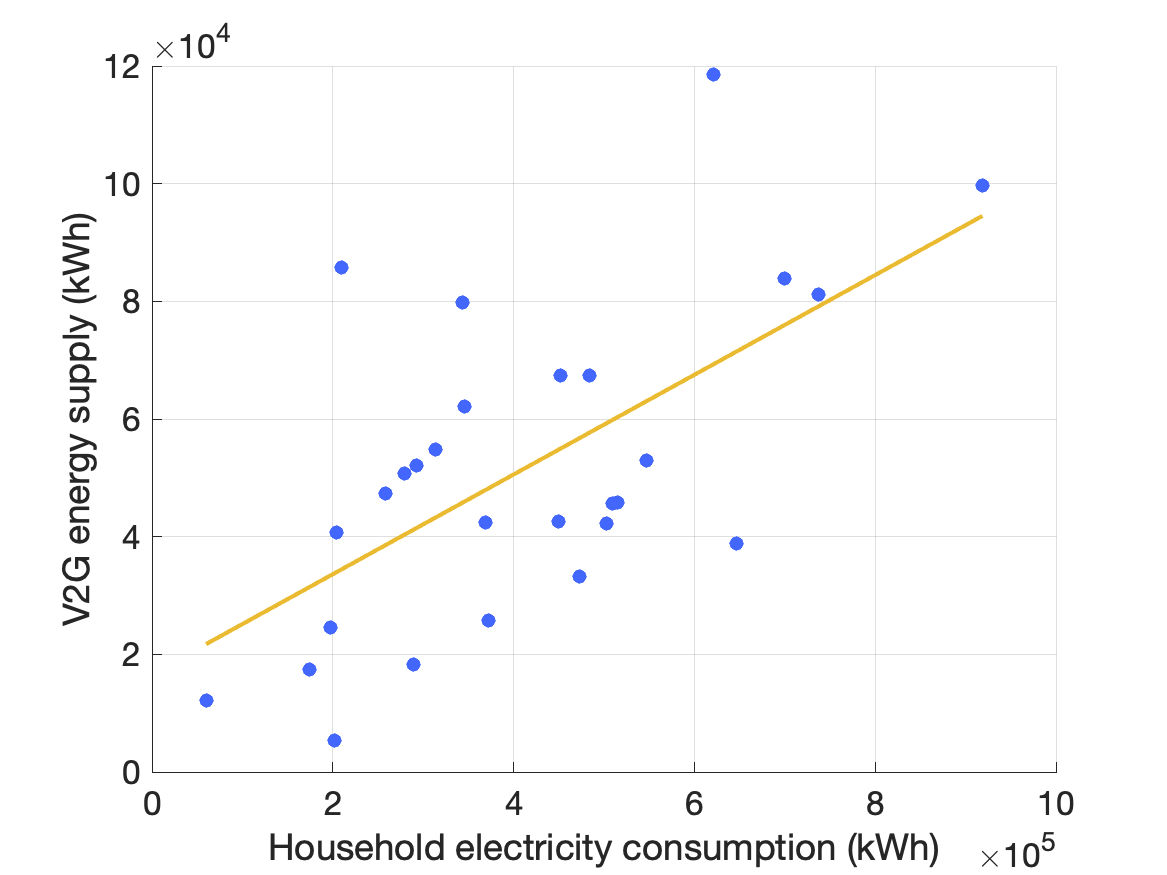}}
	\caption{V2G  energy supply versus household electricity consumption in each area during the night-time hours (midnight$-$9am and 5pm$-$midnight) of a typical day. Values correspond to a low EV penetration rate of 3\% ($\delta = 0.03$). The best fit line resulting from ordinary least squares regression (orange line) has slope 0.08 ($R^2$ = 0.38).}
	\label{fig:demsup}
\end{figure}

\begin{figure}[t!]
	\centerline{\includegraphics[width=9cm]{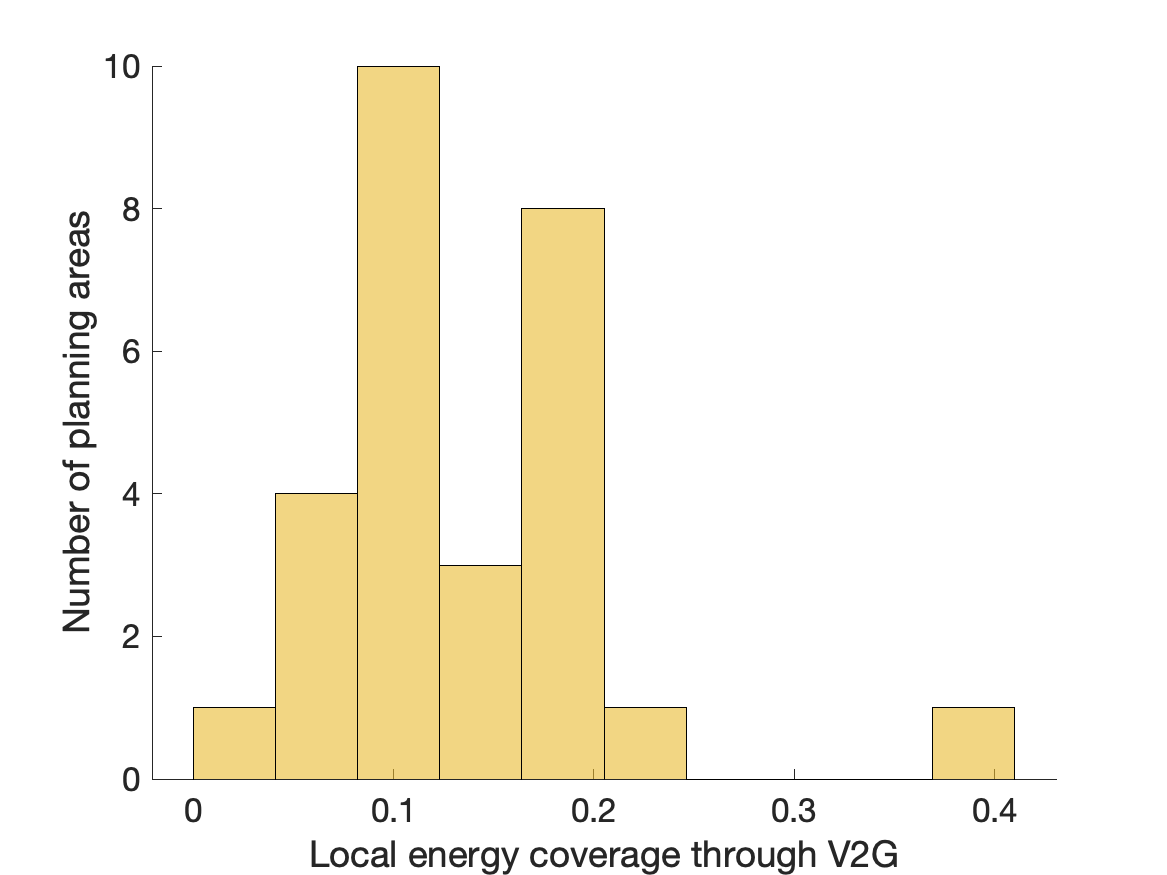}}
	\caption{Histogram of the household energy coverage by V2G in each planning area. Values correspond to a low EV penetration rate of 3\% ($\delta = 0.03$).}
	\label{fig:hist}
\end{figure}

To compare the daily energy supply from V2G (Fig.~\ref{fig:v2gmap}) with the local energy consumption, we multiplied for each planning area the average monthly household electricity consumption for September 2020 \cite{ema_2021} with the number of households \cite{mti_2015} and divided it by the number of days in the same month. Finally, by using the half-hourly system demand curve provided in \cite{ema_sdd_2021}, we estimated the fraction of the household energy consumption that is incurred during the night-time hours when PV is not available (midnight$-t_s^\text{start}$ and $t_s^\text{end}-$midnight). Fig.~\ref{fig:demsup} compares the resulting household electricity consumption, $E^\text{hh}_k$, with the V2G energy supply, $E^\text{ev}_k$.
We find a significant positive correlation ($r=0.62$, $p\text{-value}< 0.001$), which can be expected since both the aggregated household energy consumption and the number of charged EVs returning home in the evening are largely determined by the population size of each area. Nevertheless, the fluctuations around the regression line in Fig.~\ref{fig:demsup} are remarkable. We further explore the distribution of these fluctuations by calculating for each area the ratio of the night-time household electricity consumption that is covered by the V2G scheme, $E^\text{ev}_k/E^\text{hh}_k$. Fig.~\ref{fig:hist} shows the histogram of the resulting energy coverage across the different areas.
Considering the low EV penetration rate assumed for this case study ($\delta = 0.03$), the expected energy supply contribution from the basic V2G scheme is rather high, with  $E^\text{ev}_k/E^\text{hh}_k \approx 10-20\%$ for most areas. This suggests that V2G can indeed provide a significant energy storage support for the large-scale deployment of photovoltaics in cities, even if the usage of private cars is reduced in the near future \cite{duarte_2018}. The fluctuations range from $E^\text{ev}_k/E^\text{hh}_k \approx 5\%$ to $E^\text{ev}_k/E^\text{hh}_k \approx 40\%$. This again suggests that the potential of V2G for energy supply varies strongly across space and depends on the non-trivial mobility patterns of the individuals (e.g., people visiting shopping areas in the evenings before returning home as described above). Therefore, this result also confirms that the adequate planning of V2G infrastructures requires a detailed understanding of human mobility patterns (beyond home-work commuting), as is proposed in this paper.

\begin{figure}[t!]
	\centerline{\includegraphics[width=9cm]{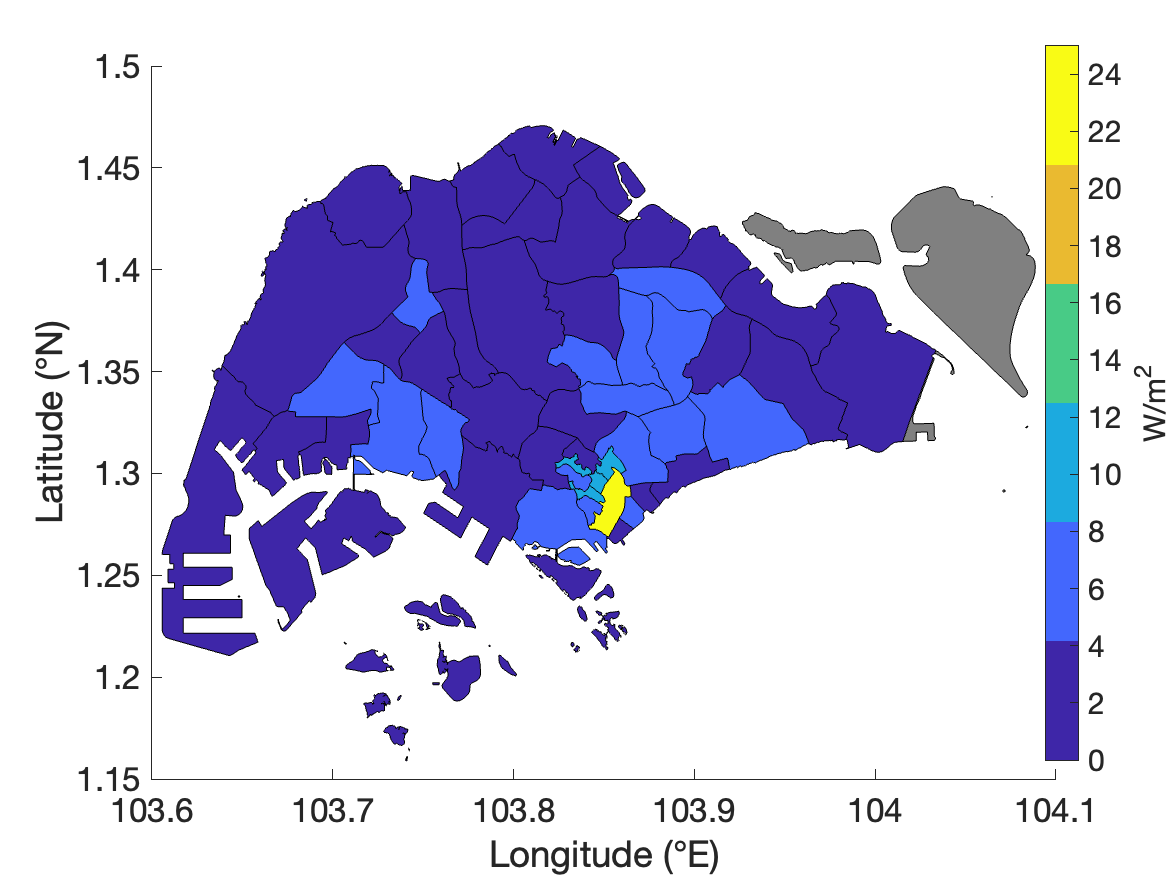}}
	\caption{Peak demand density in each planning area. Values correspond to a low EV penetration rate of 3\% 
		($\delta = 0.03$). The highest peak demand density is found in the downtown core (yellow coloured area).}
	\label{fig:pd}
\end{figure}

The peak charging demands, $P_k^\text{ev,peak}$, normalized by the square footage of each planning area, $A_k$, \mbox{$p_k^\text{ev,peak}=P_k^\text{ev,peak} / A_k$}, are depicted in Fig.~\ref{fig:pd}. The values vary substantially across the different areas. A strong `outlier` is Singapore's downtown area (`downtown core') where the maximum value is found to be $p_k^\text{ev,peak} \approx 24\,\text{W} / \text{m}^2$. To compare the peak demands to the potential electricity generation from local photovoltaics, we estimate the latter as $p_\text{pv} \approx \eta_\text{pv} a_\text{pv} I$, where $\eta_\text{pv}$ is the PV efficiency, $a_\text{pv}$ is the area of the installed PV panels relative to the total area, and $I$ is the typical irradiance. With $\eta = 0.2$ \cite{fouad_2017},  $a_\text{pv} = 0.25\,\text{m}^2 / \text{m}^2$ \cite{waibel_2021} and $I = 400 \,\text{W} / \text{m}^2$ \cite{khoo_opv_2014} this gives $p_\text{pv} = 20\,\text{W} / \text{m}^2$ for Singapore. This value is slightly lower than the maximum peak demand in the downtown core, showing that local PV alone might not be sufficient and that additional measures (e.g., smart charging schemes, additional energy supply from neighbouring areas) are needed in such high-density areas to fully cover the peak demand with solar power. The peak demand values also provide a basic guideline for determining the amount of EV charging points (parking lots with possibility for charging) that are needed in the different areas of the city. For example, assuming the model parameters given in table~\ref{table:parameters}, the highest density of charging points would be required in the downtown core, with $P_k^\text{ev,peak}/ \left(A_k P_\text{charge} \right)\! \approx \! 3800$ charging points per square kilometer. This high value clearly contradicts current efforts to reduce private car usage in cities and therefore shows the need to further extend our study with more detailed scenarios on the future mix of different transportation \mbox{modes \cite{orozco_2021}}. 

\section{Concluding Remarks}

This work has introduced the use of detailed human mobility data for the planning of V2G as a distributed energy storage system that supports the deployment of photovoltaics in cities. This framework brings together in an innovative way mobile phone data analysis and energy infrastructure engineering techniques. More specifically, anonymised mobile phone records have been used to infer the detailed mobility patterns of Singapore's population. Subsequently, this information was used to quantify how much energy from photovoltaics can be stored in EVs during the day and then released during the night-time hours within different urban areas. More extensive mobility data (e.g., higher population coverage, different observation periods) will be needed to further assess the robustness of the results presented here. Nevertheless, our initial study shows that anonymised mobile phone records can be an important means to adequately estimate the spatial and temporal distribution of the population-wide EV charging and discharging activity. This information, in turn, is an essential basis for 
the development of more intelligent (`smart') EV charging and discharging scheduling schemes that, through city-wide coordination, allow to further maximize the deployment of photovoltaics, while also reducing the local peak charging demands (especially during special events). Beyond energy storage, the insights gained from our analysis also allow to adequately explore additional potentials of V2G such as the support for local electricity distribution networks (or microgrids) with respect to voltage and frequency stability as well as network recovery after a breakdown.

In light of the current technological transition towards low-carbon cities, our work has aimed to demonstrate the  potential and some possible challenges of V2G as a means to achieve renewable-rich, integrated urban energy and transportation systems. In this respect, as ongoing work, we are bridging urban planning, urban analytics, transportation and energy infrastructure engineering, so as to investigate in more detail different future urban mobility scenarios (e.g., different forms of multimodal mobility) and their impact on the \mbox{potential of V2G}.

\section*{Acknowledgment}
The authors thank Arno Schlueter and Christoph Waibel for helpful discussions, and acknowledge CITYDATA.ai for providing the mobility data. 

\bibliographystyle{IEEEtran}
\bibliography{IEEEabrv,references}

\end{document}